# An Experimental Evaluation and Investigation of Waves of Misery in R-trees

Lu Xing
Purdue University, West Lafayette, IN
xingl@purdue.edu

Eric Lee
Purdue University, West Lafayette, IN
lee3388@purdue.edu

Tong An
Purdue University, West Lafayette, IN
an82@purdue.edu

Bo-Cheng Chu
Purdue University, West Lafayette, IN
chu206@purdue.edu

Ahmed Mahmood
Purdue University, West Lafayette, IN
amahmoo@cs.purdue.edu

Ahmed M. Aly
Facebook, Menlo Park, CA
aaly@fb.com

Jianguo Wang
Purdue University, West Lafayette, IN
csjgwang@purdue.edu

Walid G. Aref
Purdue University, West Lafayette, IN
Alexandria University-Egypt
aref@purdue.edu

## ABSTRACT

*Waves of misery* is a phenomenon where spikes of many node splits occur over short periods of time in tree indexes. Waves of misery negatively affect the performance of tree indexes in insertion-heavy workloads. Waves of misery have been first observed in the context of the B-tree, where these waves cause unpredictable index performance. In particular, the performance of search and index-update operations deteriorate when a wave of misery takes place, but is more predictable between the waves. This paper investigates the presence or lack of waves of misery in several R-tree variants, and studies the extent of which these waves impact the performance of each variant. Interestingly, although having poorer query performance, the Linear and Quadratic R-trees are found to be more resilient to waves of misery than both the Hilbert and R*-trees. This paper presents several techniques to reduce the impact in performance of the waves of misery for the Hilbert and R*-trees. One way to eliminate waves of misery is to force node splits to take place at regular times before nodes become full to achieve deterministic performance. The other way is that upon splitting a node, do not split it evenly but rather at different node utilization factors. This allows leaf nodes not to fill at the same pace. We study the impact of two new techniques to mitigate waves of misery after the tree index has been constructed, namely Regular Elective Splits (RES, for short) and Unequal Random Splits (URS, for short). Our experimental investigation highlights the trade-offs in performance of the introduced techniques and the pros and cons of each technique.

## 1 INTRODUCTION

The R-tree [10] is a commonly used spatial index. It is a balanced tree index that grows dynamically as data gets inserted into the index. When the R-tree nodes become overfull due to insertions, they split. Generally, node splits are costly, and may involve multiple adjustments and restructuring of the index at various degrees. However, because node splits are not frequent, their costs get amortized over the many index inserts that take place over time.

Some applications that utilize R-trees are insertion-heavy, i.e., these applications experience a large number of inserts that continuously occur over time. With the advent of IoT devices, more locations are being recorded and inserted into spatial databases, which is also the case for satellite data where points are collected and inserted.

In the context of the B-tree, Glombiewski, Seeger, and Graefe [8] have discovered the *waves of misery* phenomenon, where in insertion-heavy scenarios, the B-tree experiences cluttered splits of tree nodes over time that cause contention in the buffer pool, and results in unpredictable query performance [8] (Refer to Figure 1). The reason for naming this phenomenon a *wave of misery* is due to the negative impact in performance of these large numbers of tree node splits that take place in spikes over short periods of time.

In the context of the R-tree when handling insertion-heavy workloads, the questions that arise are: (1) Do R-trees experience waves of misery similar to the ones experienced by the B-tree? (2) And if the answer is yes, how can we mitigate these waves of misery? (3) Will the techniques proposed for handling the waves of misery for the B-tree work for the R-tree? (4) Do the various types of R-trees, e.g., the Linear R-tree [10], the Hilbert R-tree [11], the R*-tree [3], the RR*-tree [5] differ in how they experience waves of misery? and (5) whether the severity of the waves of misery differs with the various types of the R-trees? In this paper, we seek answers to these questions.

Figures 2(a)-(g) give the number of leaf node splits in seven R-tree variants. Figure 2(h) gives the number of non-leaf node splits in the Hilbert R-tree. We show the non-leaf node splits for the

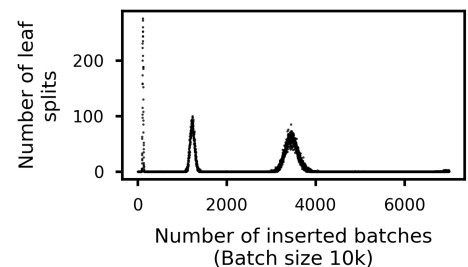

**Figure 1: Waves of misery in the B-tree. The *x*-axis is the number of batches inserted into the B-tree over time with each batch having 10,000 inserts; the *y*-axis is the number of leaf node splits in one batch.**

Hilbert R-tree only as it has the largest number of non-leaf node splits among the other R-tree variants. From the figure, the plotted counts of tree node splits resemble the shape of a wave, including leaf and non-leaf node splits. In this paper, since the leaf node splits occur much more often than the non-leaf node splits, we only focus on leaf node splits when studying the *waves of misery* phenomenon.

In this paper, we investigate the implications of the waves of misery on several disk-based R-tree variants, namely the Linear and Quadratic R-trees [10], the R*-tree [3] with and without reinsertion, the Revised R*-tree (RR*-tree) [5], and the Hilbert R-tree [11] with and without deferred split that have common yet unique features in terms of waves of misery. Experiments indicate that node utilization and range query performance are affected by the waves of misery, and that performance is stable and predictable only during the times when waves of misery do not take place. We study the impact in performance of the waves of misery under various parameter settings including the bulk-loading strategy, the initial index size, the initial node utilization, the data distribution and the page size. We show that tuning these parameters cannot resolve the impact of the waves of misery in the R-tree, which is also demonstrated to be of no use in the B-tree [8]. However, these experiments give insight on how the waves of misery can be alleviated.

In this paper, we investigate two categories of approaches to eliminate the negative effects in performance of the waves of misery, namely techniques that take place (1) during the bulk-loading phase of the index, and (2) during the batch-insertion phase. In the first category, we construct an R-tree with a variety of leaf-node fill factors. In the second category, we use different node splitting strategies including electively splitting some leaf nodes even before they get entirely full and splitting nodes at different node utilization factors. Two *Category 1* methods are proposed in [8]. In this paper, we propose two *Category 2* methods. We compare and contrast all four methods. Their impact on query performance and on mitigating the effects of the waves of misery are investigated.

Our experiments reveal some interesting findings. It is surprising that the severity of the waves of misery differs among the R-tree variants. Traditionally, it is well known that the Linear and the Quadratic R-trees are outperformed by the other R-tree variants in terms of query performance [3]. Despite that, our experiments demonstrate that the Linear and Quadratic R-trees experience waves of misery to a much lesser extent than the Hilbert R-tree. This performance of the Linear and Quadratic R-trees has provided insight on how to reduce the effect of the waves of misery in the context of the other trees, e.g., the Hilbert R-tree. We provide several heuristics for adoption within the insertion and split algorithms of the Hilbert R-tree. These heuristics help diminish the effect of waves of misery in the Hilbert R-tree without sacrificing much of the Hilbert R-tree's superior query performance.

The contributions of this paper are as follows:
- This paper is the first to study and analyze waves of misery in the R-tree and its variants. We demonstrate that the different variants of disk-based R-trees suffer in terms of performance from waves of misery at various degrees. We investigate the reasons for this problem.
- We introduce two new methods to mitigate waves of misery for variants of the R-tree that suffer from the waves of misery the most.
- We conduct thorough experiments using synthetic and real data sets. We study the performance of the new mitigation methods that we introduce in this paper for dealing with the waves of misery for the R-tree variants. We contrast these new methods with the existing mitigation methods that have been developed originally for the B-tree and that we adapt for the R-tree. We highlight the trade-offs in performance of each of the methods as well as their pros and cons, and provide recommendations towards the end of the paper.

The rest of the paper is organized as follows. Section 2 discusses the related work. Section 3 presents the experimental setting for our investigation. Section 4 demonstrates the waves of misery phenomenon in several variants of the R-tree. Section 6 presents several measures for assessing the severity of the waves of misery. Section 7 presents techniques to mitigate and alleviate the waves of misery while Section 8 presents the performance results of all the proposed techniques. Finally, Section 9 concludes the paper as well as summarizing trade-offs for each tree and remedy, and provides some recommendations.

## 2 RELATED WORK

Glombiewski et al. [8] are the first to discover the waves of misery. Their experiments focus on the B-tree. They assess the waves of misery for the B-tree under various conditions. They propose two remedies to alleviate waves of misery for the B-tree, both are during the bulk-loading phase.

The R-tree [10] is a dynamic index suited for multi-dimensional data objects. Searching the R-tree starts from the root and descends the tree. Guttman [10] has proposed three methods to split an R-tree node: (1) the exhaustive method, (2) the quadratic method, and (3) the linear method. In the quadratic algorithm, two seed objects are first identified. For each of the remaining objects, compute the difference of the increase in area of the covering rectangles if that object were to be added into each of the two partitions. The object with the largest difference is chosen and is assigned to the partition with the smaller rectangle enlargement. In the Linear R-tree algorithm, two seed objects are identified, and the remaining objects are assigned to each partition with minimum increase in the area of the bounding rectangle.

One drawback of the above algorithms is their lack of optimizing the size of the overlapping regions among the split partitions. Thus, if a query range intersects any of these overlapping regions, multiple subtrees have to be descended to answer the query. Algorithms have been proposed to eliminate, or at least reduce, the overlap. New R-tree variants and their corresponding algorithms have been introduced, e.g., [1, 3, 6, 9, 11].

The R*-tree adopts a combined optimization that minimizes both the areas and the overlap between the enclosing rectangles [3]. Upon a split, the R*-tree generates several candidate distributions, and computes three goodness measures; the area, the margin, and the overlap. By optimizing these measures, the R*-tree reduces the number of paths to traverse during a search. The R*-tree defers a node split by reinserting the object farthest from the center of the bounding rectangle into some other node in the same tree level.

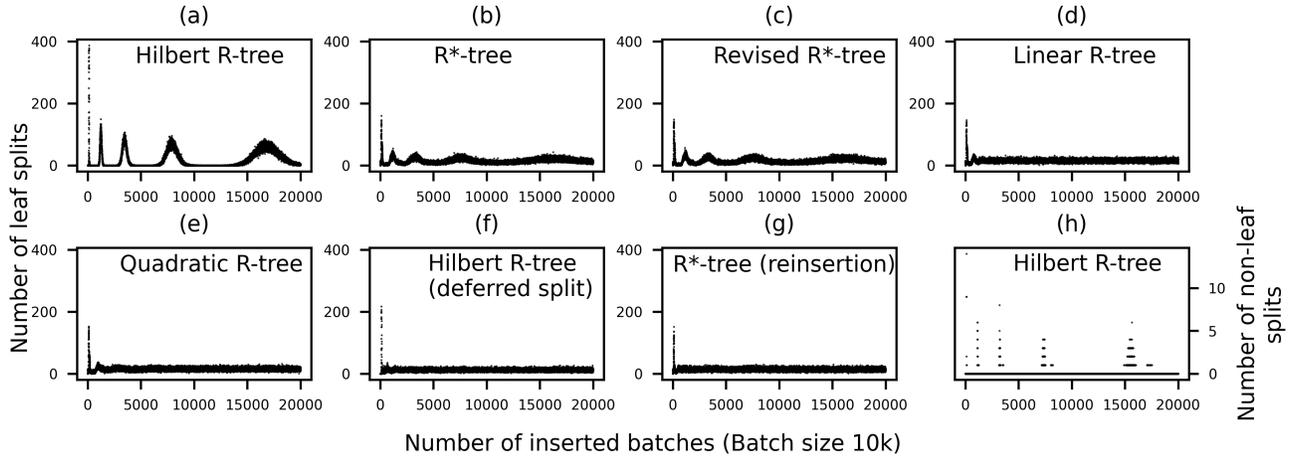

Figure 2: Demonstration of the waves of misery in seven variants of the R-tree

The Hilbert R-tree is a variant of the R-tree that imposes a linear ordering on the data rectangles stored in the R-tree [11]. The Hilbert value of the center of each data rectangle is computed, and thus, a linear ordering of the rectangles is obtained. The bounding rectangles are stored in the index nodes. The Hilbert R-tree allows for deferred splitting. For an $s$-to-$(s + 1)$ splitting policy, the overflowing node tries to push some of its entries to one of its $s − 1$ siblings. If all of them are full, then a new node is created, and the $s$ nodes are split into $s + 1$ nodes. This splitting policy drives the utilization of the Hilbert R-tree to be as close as 100% [11]. In our experiments, we use the '1-to-2' and '2-to-3' split policies.

Ang and Tan present a linear algorithm for minimizing region overlaps among R-tree nodes [1]. In an overflowing node, all objects are distributed into two separate lists per dimension. The dimension that offers a more even distribution is used to split the node. If there is a tie, the dimension with the smallest overlap is favored. If another tie occurs, the dimension with the smallest total coverage is chosen.

Korotkov [14] proposes a double sorting-based node-splitting algorithm. In the one-dimensional case, two arrays are used to store the objects based on the lower and the upper bounds separately. Arrays are sorted to choose two objects with minimum overlap as seeds. Each seed represents a group. Then, unambiguous objects are distributed to each group first. The remaining objects are sorted according to the interval centers. Then, the objects are distributed such that the total number of entries in the first group is $m$ ($m$ is the minimum number constraint), and the remaining entries go to the second group.

The Revised R*-tree (RR*-tree) [5] enhances over the R*-tree in several ways. The reinsertion policy of the R*-tree is abandoned. Also, the R*-tree has a relatively expensive overlap optimization that is only performed in the lowest non-leaf level (the one above the leaf nodes). With a redesigned algorithm, overlap optimization can be applied to all non-leaf levels in the RR*-tree. Also, the balance of the splitting pages is added as an optimization criterion. Another improvement is on high-dimensional data. Since it is possible that bounding boxes with zero volume occur when using high-dimensional data, the volume-based optimization becomes less effective. In such situations, a lower dimension perimeter-based optimization is used.

There are several techniques for bulk-loading the R-tree. One bulk-loading method is sort based [12]. Objects are ordered based on some criterion, e.g., based on the Hilbert value of the centroid of the object, then are packed into pages according to that order. Another bulk-loading technique is the Sort-Tile-Recursive (STR, for short) [15], where objects are sorted into tiles recursively using one dimension at a time. For example, with two-dimensional data, objects are first sorted by their $x$-coordinates and are partitioned into vertical slices. Then, in each slice, objects are sorted by their $y$-coordinates and are grouped into runs of length $b$ ($b$ is the maximum number of objects that one node can hold) into one node.

## 3 EXPERIMENTAL SETUP

We use a 192 core machine having Intel(R) Xeon(R) Platinum 8168 CPU @ 2.70GHz with a 2 TB memory running Linux kernel 4.15.0. We implement all the proposed methods within the Java indexing library XXL [17]. We use the following synthetic and real two-dimensional point data sets:

**Synthetic Data:** We use three two-dimensional synthetic data sets in the experiments. All synthetic data sets are in the unit space $[0, 1] \times [0, 1]$. The first point data set is uniformly distributed. The $x$ and the $y$ coordinates of the points are generated uniformly in the space. The second point data set is normally distributed. Both the $x$ and $y$ coordinates of the points follow normal distribution with Mean 0.5 and Standard Deviation $\frac{1}{6}$. The third data set is an artificial distribution *Bit* described in [4]. *Bit* is a point distribution with floating point vectors generated as follows: each bit of the mantissa of their components is set with probability 0.3. The data is generated via SpiderWeb [13] inside the unit space.

**Real Data:** We use the following two real data sets in the experiments: OpenStreetMap [16] and Twitter data set. We download the US Midwest map from OpenStreetMap, and extract the coordinates of points of interest from the map. The Twitter data contains tweets that have been gathered over the period from 2013 to 2014. We use the data that are collected in 2014/10/12. Only the tweets that have spatial coordinates inside the US are considered. The format of each tweet is: tweet identifier, tweet timestamp, tweet location (longitude and latitude coordinates), and the text body of the tweet. Both data sets are shuffled before use.

All the experiments are performed using the following R-trees: the Linear R-tree, the Quadratic R-tree, the R*-tree with and without reinsertion (if not indicated, R*-tree means no reinsertion in this paper), the Revised R*-tree (the RR*-tree), and the Hilbert R-tree with and without deferred splits (2-to-3 splits). All the indexes are stored as files on disk. There is a buffer in memory with a fixed size, which is the size of the available buffer in terms of the number of nodes. The buffer is able to hold 1000 nodes unless stated otherwise. For the rest of the paper, we assume that exactly one node fits per disk page, and hereafter we use the two terms: node and page interchangeably. All index nodes are fixed in the buffer. Only leaf nodes can be fetched and flushed to disk. All the R-trees are bulk-loaded with points sorted based on the Hilbert value. Points are inserted one at a time after bulk-loading. The statistics are collected per batch with each batch consisting of 10,000 points. The initial index size is 10 million points unless stated otherwise. All the leaf nodes are packed with an initial utilization of 90% unless stated otherwise. The page size is 16KB unless stated otherwise.

We use three kinds of queries for the different data sets:
**Uniform queries:** The center of each query square follows a uniform distribution in the unit space. The width of the query square is determined by the query area. These uniformly distributed queries are used on the synthetic uniform data with sizes ranging from 0.0001%, 0.001%, 0.01%, 0.1%, 1%, and 5% of the entire space.
**Nonuniform queries:** We simulate nonuniform queries by first fixing focal points at random, and then generating query centers following a normal distribution around the focal points. Let $r$ be the closest distance from a focal point to the unit space boundaries. In order to ensure that all queries reside inside the unit space, the standard deviation ($\sigma$) of the normal distribution is chosen such that 99.7% of data lies inside the universe ($3\sigma \leq r$). The closer to the central point, the higher the probability of the query centers, thus the higher the density of the queries.
**Queries with fixed selectivity:** This type of queries is particularly useful for real data sets. Because real data is non-uniformly distributed, we want to guarantee that a test query returns a certain number of objects, i.e., we need to be in control of the selectivities of the test queries. In order to do that, one point is randomly chosen in the underlying space, and a certain number of neighbors are retrieved using the $k$-nearest neighbours query with $k$ being the target selectivity. The query rectangle is formed as the minimum rectangle holding all the neighbors. It may happen that when we form the minimum bounding rectangle around the target number of objects that the resulting rectangle ends up containing more objects than the targeted selectivity. We discard this range query if the actual query result is more than double the selectivity. We apply queries with fixed selectivities ranging from 0.0001%, 0.001%, 0.01%, 0.1% of the number of objects in the newly updated data set after each batch.

100 queries are issued after each batch insertion. I/O access is measured by the number of buffer evictions as the buffer is filled up with tree nodes. I/O latency, i.e., the total elapsed time by I/O operations, is used to compare across the various R-tree variants.

## 4 PROBLEM DEFINITION

First, we show what the waves of misery are. In a write-intensive spatial database, e.g., satellite data and IoT devices' locations being inserted into R-tree indexes, we simulate this scenario by inserting batches of data over time into the R-tree, where each batch contains 10,000 points. We count the number of splits that happen during one batch, and plot all the batches in Figure 2. Seven different R-trees are used for comparison. Figures 2a-g give the number of leaf node splits. Figure 2h gives the number of non-leaf node splits in the Hilbert R-tree.

From Figure 2, all R-trees have spikes (waves) of leaf node splits that are cluttered over short periods of time. We refer to these spikes as *waves of misery* as they affect the performance of the index during the wave. The Hilbert R-tree (Figure 2a) has the most obvious waves, and hence is affected the most by the waves of misery. The first wave is the highest. The following waves show a decrease in height over time but with an increase in width. The number of splits is almost zero in-between the waves.

The R*-tree without reinsertion (Figure 2b) and the RR*-tree (Figure 2c) also have waves of misery, and they are almost identical, both in the heights of the peaks and the batch numbers of each peak. All of their waves are smaller than those of the Hilbert R-tree. In between the waves, the number of splits is nonzero. Since one of the improvements in the RR*-tree is on high-dimensional data and we only use two-dimensional data, the difference between the RR*-tree and the R*-tree (without reinsertion) is not much. We only show the result of the R*-tree in later experiments.

The Linear R-tree (Figure 2d) and the Quadratic R-tree (Figure 2e) are similar to each other, and both of them do not exhibit significant waves of misery except for their first waves. The numbers of splits become stable after the second wave.

For the Hilbert R-tree with deferred split (Figure 2f), waves of misery are eliminated effectively except the first two waves. This is expected because nodes do not have to split when they overflow, rather they can push the overflow objects into one of the siblings of the overflowing node. However, this results in longer insertion time as illustrated in Figure 3a (left). The points are plotted at an interval of 100 batches. The excess time in the deferred split policy is spent to find a sibling node with some empty space as well as adjust the maximum bounding rectangles for both nodes. The R*-tree with reinsertion is another example of deferred split and has longer insertion times as in Figure 3a right.

When waves of misery take place, they put the buffer manager in contention. Next, we investigate whether buffer utilization and eviction have effect on the presence or lack of waves of misery. We use two buffer sizes of 1000 and 5000 pages, and show the results for the Hilbert R-tree in Figure 3b. Buffer utilization is calculated as the quotient between the used buffer pool bytes and the total available buffer pool bytes. The plot declines after each wave and accumulates in between waves. Buffer eviction is counted as the number of evicted buffer slots per batch. There is a sharp peak or sharp turning point at each wave, especially around Batch 100 for the buffer of size 1000 pages.

We also use Data Set *Bit* [13] and two real-world data sets: OpenStreetMap and Twitter. The waves of misery of each R-tree (Figure 3c) resemble their counterparts of the uniform data sets.

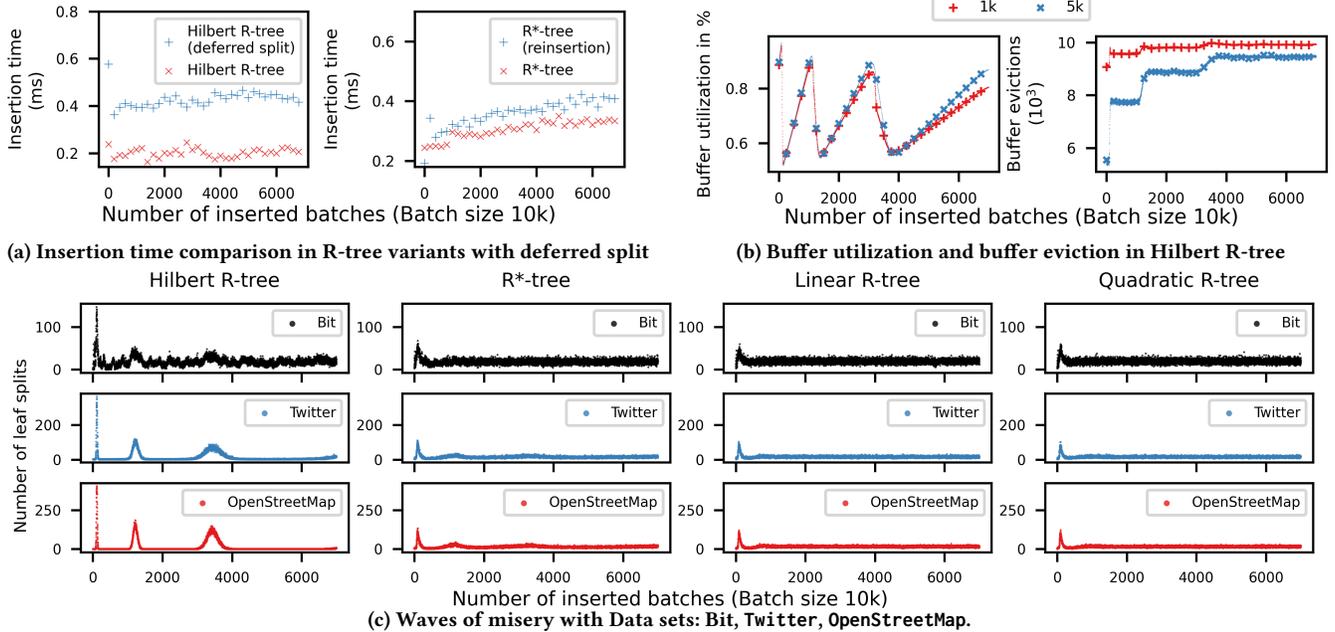

Figure 3: (a) Insertion time comparison (points are plotted at an interval of 100 batches), (b) Buffer contention, and (c) Waves of misery for Data sets: Bit, `Twitter`, and `OpenStreetMap`.

Waves of misery are most severe in the case of the Hilbert R-tree, followed by the R*-tree, and then the Linear or the Quadratic R-trees. The least is the Hilbert R-tree with deferred splits and the R*-tree with reinsertion. This provides insight on how to mitigate waves of misery as will be discussed in Section 7.

## 5 IMPACT OF WAVES OF MISERY

### 5.1 Query Performance

We perform range search queries of various range sizes, and plot the number of I/Os in Figure 4a. We also issue fixed selectivity queries on `OpenStreetMap` in Figure 4b and nonuniform queries on uniform data in Figure 4c. We use the mean of a moving window (runmean) equal to 100 batches to plot the number of I/O accesses. The I/O latency and the number of I/O accesses of the 90th percentile (tail latency) are given in Figure 4d first two rows. The three kinds of queries show consistent results. The query performance of the Hilbert R-tree shows a staircase pattern of increase. Each abrupt increase in I/O latency matches the end of one wave of misery. This demonstrates that waves of misery are the cause of the abrupt increase in the number of I/O accesses. Notice that this abrupt degradation in query performance is not due to the increase in the depth of the tree because all the index (non-leaf) nodes are pinned inside the buffer. As waves of misery make buffer management in contention, large amount of leaf nodes are evicted from the buffer. Thus, querying has to fetch the leaf nodes from disk again. The increase in the number of I/O accesses in the case of the R*-tree is not smooth either. Similar to the Hilbert R-tree, the increase in the number of I/O accesses matches the end of the wave of misery where too many splits take place. Notice that Linear and Quadratic R-trees have almost no waves of misery after 2000 batches. As a result, the number of I/O accesses for both the Linear and Quadratic R-trees increases smoothly and not as abruptly as in the cases of the Hilbert R-trees.

### 5.2 Node Utilization

In Figure 4d Row 3, we examine the average utilization of leaf nodes after each batch. During bulk-loading, leaf node utilization is fixed at 90% except the last leaf node. Leaf node utilization increases in the first few batches. When the first wave of misery happens, most leaf nodes are full and they split almost at the same time, resulting in a drop in node utilization. With plenty of free space in leaf nodes, utilization accumulates as objects are inserted until the second wave occurs, and the utilization drops again. The Hilbert R-tree has the largest oscillations followed by the R*-tree. Both the Linear and Quadratic R-trees have nearly stable node utilization after Batch 2000. The Hilbert R-tree has the highest and the lowest transient utilization. This result is consistent with the waves.

### 5.3 Number of Leaf Nodes and R-Tree Height

In Figure 4d Row 4, the number of Hilbert R-tree leaf nodes grows in a staircase pattern. There are batches that do not have any leaf node increase. In between the plateau area, there is an extensive growth in leaf nodes, while the remaining three R-trees have steady growth in the number of leaf nodes with data insertions. We plot the tree height after each batch. The increase in height is rare.

## 6 ANALYSIS

For the four types of R-trees, we investigate what factors affect the waves of misery including the bulk-loading strategy, the initial index size after bulk-loading, the initial utilization factor at bulk-loading time, the data distribution, and the page size.

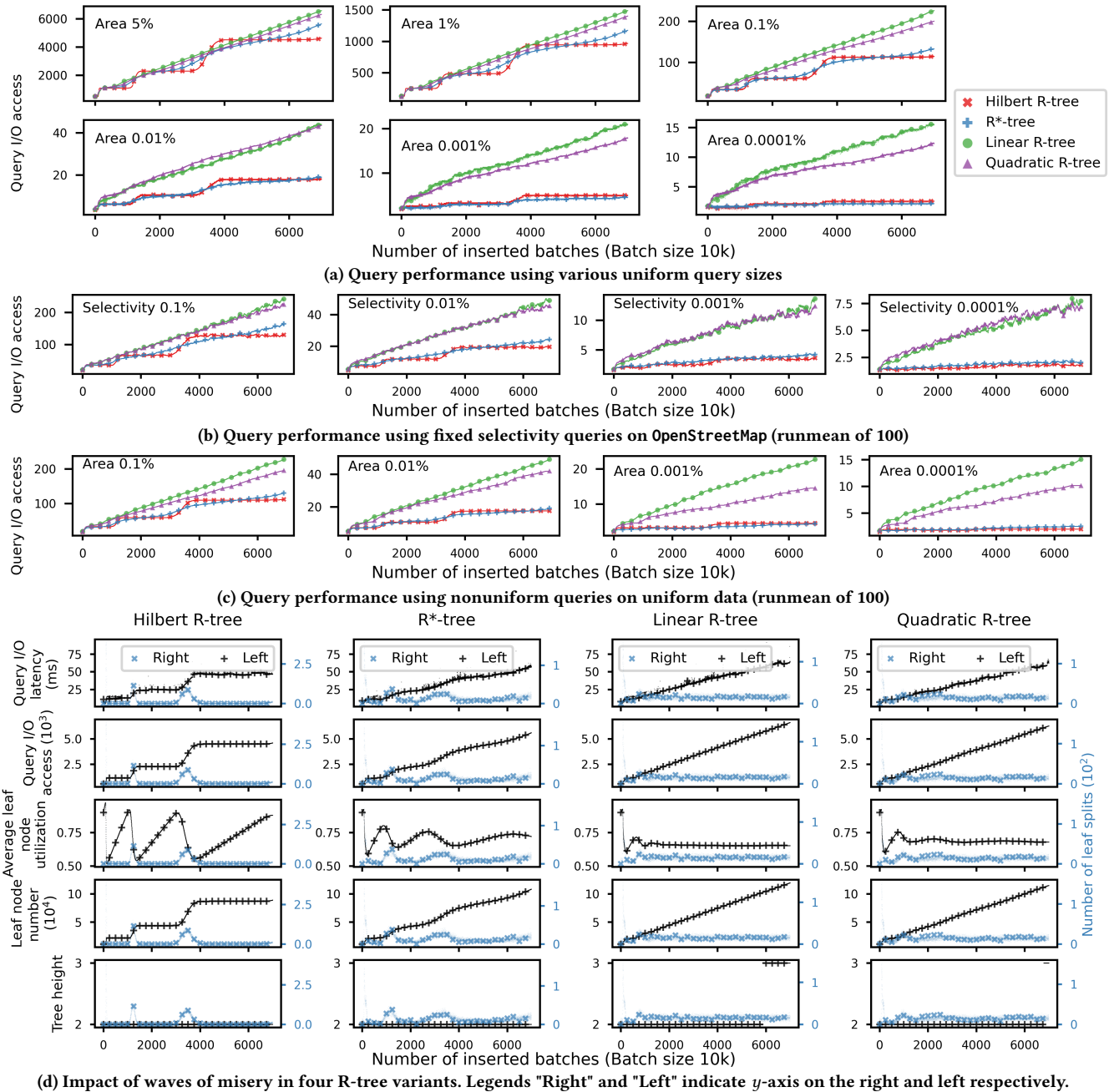

Figure 4: Query performance of uniform, nonuniform and fixed selectivity queries and impacts of waves of misery

**Effect of Bulk-loading Strategy:** There are various bulk-loading strategies, including the Sort-Tile Recursive (STR) [15] and sort-based strategy [12]. We bulk-load the R-trees using STR and plot their waves of misery in Figure 5a. Since the Hilbert R-tree sorts the objects based on their Hilbert values, rather than their $x$ and $y$ coordinates (as performed in STR), we exclude the Hilbert R-tree from this experiment. The highest wave is more than 200 splits in all the four R-trees of Figure 5a, which are greater than those in Figure 2. Even for the R*-tree (with reinsertion), the Linear R-tree, and the Quadratic R-tree that originally do not exhibit any obvious waves of misery after their first waves, show oscillating waves after Batch 1000 as in Figure 5a. As can be seen from the figure, the problem of the waves of misery is more severe in the case of the STR strategy.

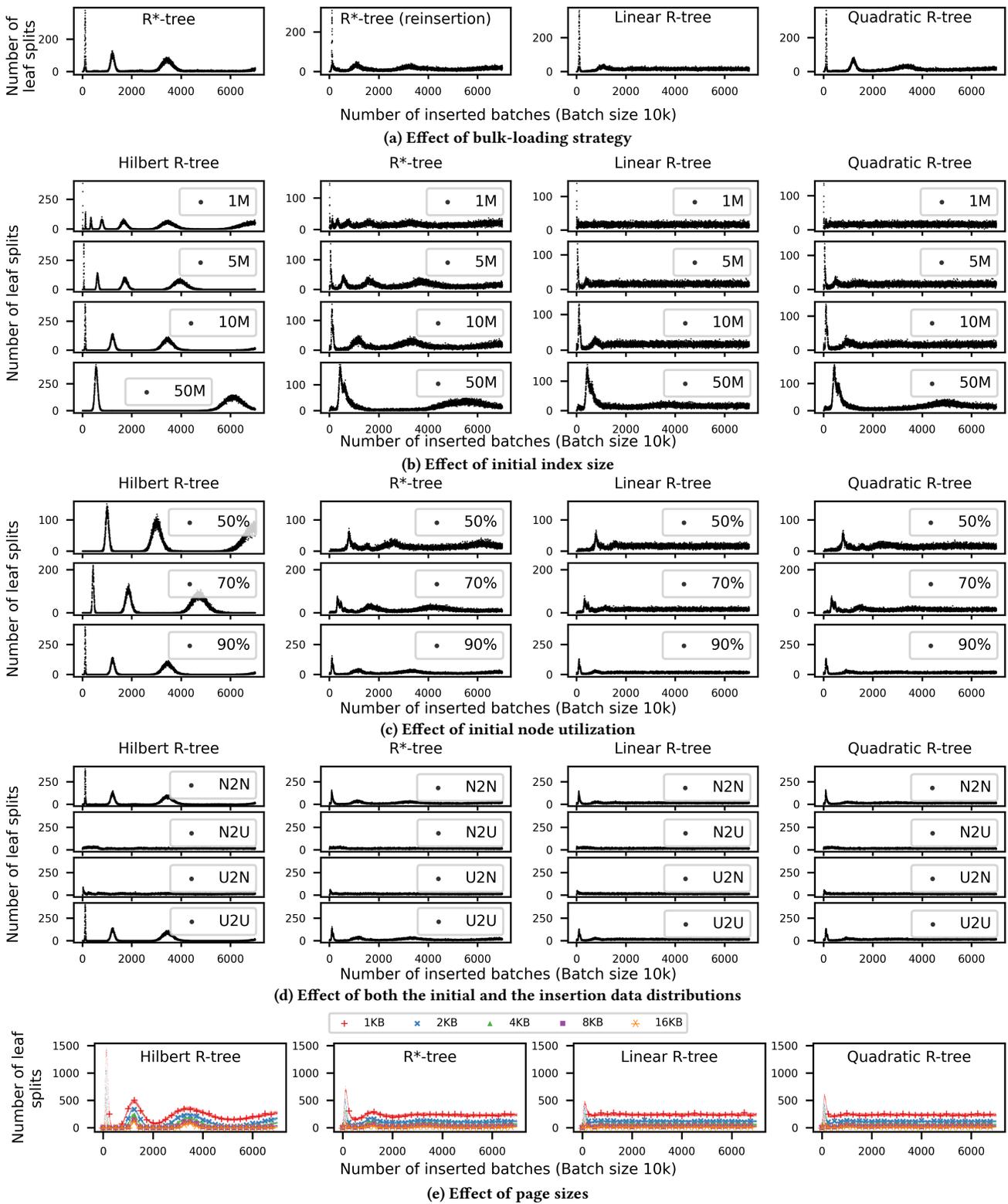

Figure 5: Analysis on the waves of misery

**Effect of the Initial Index Size:** The motivation behind focusing on the initial index size at bulk-loading time is the following. *Hypothetically, if we know the ultimate and final index size at index construction time, we would allocate the needed number of leaf nodes in advance, and hence eliminate the waves of misery as no splits would take place.* This makes the initial index size a sensitive parameter. We compare four R-tree variants under four initial sizes: 1M, 5M, 10M, and 50M points. In Figure 5b, all four index sizes still show waves of misery during batch inserts. However, the larger the index, the more batches it needs to show a following wave. The widths of the waves are also different. The larger the index, the wider each wave. This behavior is consistent in all R-tree variants. While a good observation, only changing the initial index size does not stop waves of misery from happening.

**Effect of Initial Node Utilization at Bulk-loading Time:** If some free space is allocated at bulk-loading time, short-term insertions can be ingested without node splits. However, this would not eliminate the waves of misery altogether. In Figure 5c, we compare the split counts under initial node utilization of 50%, 70%, and 90%. The 90% utilization tree needs the least number of insertions to have its first wave. The amplitudes of the waves differ as well. The waves are more compact in the trees with higher utilization ratios. *Thus, tuning the initial utilization does not eliminate the waves, but rather delays and weakens them.*

**Effect of the Data Distributions of the Initial and Batch Phases:** The previous experiments use uniform data both for the construction and insertion phases. Here, we experiment with data that follow normal distribution in both phases. There are four combinations: Normal/Normal (N2N), Uniform/Uniform (U2U), Uniform/Normal (U2N), and Normal/Uniform (N2U). Similar to [8], N2N and U2U (Figure 5d 1st and 4th rows) show the most obvious waves of misery. If the distributions do not match, although there are oscillations before batch number 800, waves are still negligible.

**Effect of Page Size:** We compare four R-tree variants with different page sizes: 1KB, 2KB, 4KB, 8KB, and 16KB in Figure 5e. With a 1KB page size, the peaks are the largest and the valleys in between the waves also have the largest number of splits. The 2KB page size forms waves of misery beneath those of the 1KB page size, with the same oscillating pattern and frequency, but with a smaller number of splits. The waves of misery for the 4KB, 8KB, and 16KB page sizes have even lower peaks. When the initial index size is kept the same, a smaller page size leads to more leaf nodes, thus the wave is higher. This result is consistent for all four R-trees.

To summarize, switching to a different bulk-loading strategy, or tuning the initial index size, the initial node utilization or varying the page size do not eliminate waves of misery, but rather delay them. Although waves of misery are weaker if data distributions do not match between the initial bulk-loading and the batch insertion phases of data upload inserted the R-tree, these data distributions are usually unknown beforehand. However, the unmatched distributions give insight into how to mitigate waves of misery.

## 7 MITIGATING WAVES OF MISERY

In [8], new algorithms at the bulk-loading phase are proposed to have diverse node utilization factors before the batch insertion phase starts. We introduce these algorithms here to test them in the context of the R-tree. Then, we propose two new methods that take place during the batch insertion phase. One main advantage of these new methods is that a rebuild of the R-tree is not required.

Another observation is that there are R-trees that show negligible waves of misery, mainly, the Linear and Quadratic R-trees, and the R-tree with deferred splits. In Section 7.3, we present in-depth analysis of how the split algorithms for the Linear and Quadratic R-trees affect the waves of misery.

### 7.1 The Bulk-loading Phase

*7.1.1 Sound Remedy (SR).* Sound Remedy [8] has been introduced to mitigate waves of misery for the B-tree. The basic idea is to keep the leaf node utilization in balance (steady state) when the tree is constructed, and maintain this balance at all times. This prevents time-clustered leaf node splits from happening.

Assume that a node, say $Q$, is at least 50% full. Let $j$ be the number of objects in $Q$, i.e., $(B + 1)/2 \leq j \leq B$, where $B$ is the maximum number of objects that a node can hold. The steady-state probability $P_j$, that a leaf node has $j$ objects, is $P_j = \frac{B}{j(j+1)}$, and the sum of the probabilities across all leaf nodes is: $\Sigma_{j=(B+1)/2}^{B} P_j = 1$. During bulk-loading, to start from a steady-state, $j$ is randomly decided by $P_j$. After $j$ is determined, the next $j$ objects are put into one node and no more objects are added. When the remaining objects are less than $(3B + 1)/2$, we switch to a final round: If there are at most $B$ objects, they are put in one leaf node. Else, the objects are distributed equally over two leaf nodes. This algorithm guarantees that all the leaf nodes are at least half full and the probability that a record insertion triggers a split of a leaf node is constant. The proof follows the basic ideas of fringe analysis [2, 7, 18] and is omitted.

*7.1.2 Linear and Random Practical Remedies (LPR and RPR).* There are other methods to determine node utilization factors, namely practical remedy. The linear and random practical remedies (LPR and RPR respectively) [8] are illustrated below by an example. If the bulk-loading target utilization is 90%, LPR assigns the first node to be 100%, the second to 80%, the third to 99%, the fourth to 81%, etc. The value is chosen in this alternating way. However, RPR requires a target range as well. With 90% target utilization and 10% target range, RPR chooses a random number from 80% to 100%. With a smaller 5% range, the random number is constrained within 85% to 95%. One advantage of LPR and RPR over Sound Remedy (SR) is that the target utilization can be tuned. The theoretical analysis of SR guarantees that the average utilization is about 69% (ln 2).

### 7.2 Batch Insertion Phase

The previous methods apply during bulk-loading. However, in situations when rebuilding an existing R-tree is costly, mitigating waves of misery during batch insertion can be a better option.

*7.2.1 Unequal Random Split (URS).* The waves of misery happen because nodes get filled synchronously. The core idea of the above methods is to keep the filling ratio of the leaf nodes in a steady state starting from the bulk-loading phase. However, the filling ratio can also be tuned during the batch insertion phase. In Figure 2, the overflowing leaf nodes are split in half, i.e., the two resulting leaf nodes are half full. If the split is intentionally made unbalanced, the resulting two leaf nodes will receive a different number of entries, and thus simulating the effect of SR. However, if the imbalanced split is kept at a fixed ratio, e.g., 30% and 70%, waves of misery cannot be avoided either. The reason is that nodes that are 70% full are being filled at the same pace, forming one round of waves, while nodes that are 30% full are filled up together, forming another round of waves. We demonstrate the performance of unequal fixed split (UFS) in Section 8.3. With a random number chosen between 30% and 70%, the remaining node utilization becomes closer to the steady state. Thus, URS is faster than UFS in achieving stability.

*7.2.2 Regular Elective Split (RES).* One observation from the Hilbert R-tree is that there are no leaf node splits in between the waves of misery. We can force leaf nodes to split regularly and sequentially such that leaf nodes split before they actually overflow. We allow one overflow page for each leaf node in case the node gets filled before its split time. Although this may deteriorate the average leaf node utilization, the overall index becomes more deterministic especially in query performance. We give a unique increasing identifier to each leaf node at its creation time, and start the split process from the first leaf node. During insertion, a point is inserted into either a leaf node, or an overflow page if the leaf node is full, or a split is triggered if all overflow pages are full. Besides this split, a split is performed regularly every $m$ insertions on leaf nodes based on the node identifier. Parameter $m$ can be optimized based on the workload. This way, each node has a chance to get split. Thus, waves of misery can be avoided before they occur. Hence, splits will be distributed over time, and waves of misery will be flattened.

## 7.3 Effect of the Split Algorithm

Refer to Figure 2, both the Linear and Quadratic R-trees do not have discernible waves of misery as those in the Hilbert and R*-trees. Since these R-trees differ most in their splitting algorithms, we dissect the splitting process, and how the splitting algorithm affects the waves of misery. Consider the Linear R-tree's split algorithm. First, two seed points need to be found. The remaining points are distributed by computing the area of the bounding rectangle. There is no optimization in minimizing the overlap of the resulting two leaf nodes. Based on the experiments, we observe that many of the splits are skewed, that is, the smaller leaf node ends up residing almost completely inside the larger one. To confirm this, first, we define *SplitOverlapRatio* as follows:

$$\text{SplitRatio} = \frac{Area(\text{Overlap})}{Area(\text{MBR(Small leaf node)})} \quad (1)$$

If *SplitOverlapRatio* is close to 1, then the overlapping area is almost as large as the small leaf node. This implies that the smaller leaf node almost resides inside the larger one. We compute the average value of *SplitOverlapRatio* for all R-tree variants.

As expected, the Linear R-tree has the highest *SplitOverlapRatio* (95%), indicating that its split strategy results in the situation described above, i.e., on average, 95% of the area of the small leaf node is covered by the large leaf node. The lowest *SplitOverlapRatio* value is the R*-tree's (0.048%). *SplitOverlapRatio* for both the Quadratic and Hilbert R-trees is around 1/3. When a point is inserted into the Linear R-tree, a leaf node with the smallest area is chosen for insertion. Thus, if a point falls within the range of the overlapping region, it is inserted into the small leaf node. If more points are inserted into the overlap region, then the small leaf nodes will be filled faster. The above statement holds only when the overlap region has a higher chance of points being inserted into it. This means that the size of the overlap region is greater than that of the non-overlapping one. We compute *SplitRatio* as follows:

$$\text{SplitRatio} = \frac{Area(\text{MBR(Small leaf node)})}{Area(\text{MBR(Large leaf node)})} \quad (2)$$

We observe that the Linear R-tree has an average *SplitRatio* of 70% with standard deviation of 20% indicating that the smaller partition during a split occupies more than half of the larger partition.

In an area small enough like a bounding rectangle of a leaf node, we can assume data points are distributed uniformly. When there is a point, say $p$, waiting to be inserted into a leaf node, the probability of inserting $p$ into the overlap region is higher because it has a larger size. **Thus, when $p$ can be inserted into multiple leaf nodes because of the overlap, the leaf node with the smaller bounding rectangle is chosen. Thus, the smaller leaf node will be filled much faster than the bigger one in the Linear R-tree, which alleviates the waves of misery, but has poor query performance because of the larger overlap region.**

However, this does not explain why the Quadratic R-tree has negligible waves of misery because *splitOverlapRatio* is only 1/3 in the Quadratic R-tree. We trace all the leaf nodes split in Batches 200-1000, and record the type of leaf node that gets split. There are two types of leaf nodes: an original bulk-loaded leaf node, or a descendant from a previous split, i.e., a split node. Furthermore, there are two kinds of split nodes: *Large Child*, i.e., the one with a larger area, and *Small Child*, i.e., the one with a smaller area.

Within Batches 200 to 1000, 60.5% of the split nodes are *Large Child*, 35.2% are *Small Child*, and only 4.2% are *bulk-loaded nodes*. This suggests that more than half of the splits are from *Large Child*. Thus, *Large Child* fills faster during this time window. The reason is that after each split, each of the two leaf nodes has half the entries. Due to the same reason, we can assume that the points are distributed uniformly within one leaf node, and so are the inserted points. Thus, a leaf node with a larger area is more likely to have more points inserted. This explains why larger leaf nodes are split more often than the other ones in the Quadratic R-tree.

To prove this hypothesis, we record the area difference between the two leaf nodes during each split and sum the area differences across all batches.

The Linear R-tree has the largest area difference of 9.15, followed by the Quadratic R-tree (5.79), and the R*-tree (2.08). The Hilbert R-tree is the smallest among all at 0.94. Thus, with a larger area difference, the *Large Child* receives more points and gets filled sooner. Thus, the split is spread across all batches. The order of area differences is consistent with the waves of misery in Figure 2 as the Hilbert R-tree that shows the smallest area difference has the most obvious waves of misery, followed by the R*-tree, while the Linear and Quadratic R-trees that have more area differences do not exhibit obvious waves.

## 8 EVALUATION RESULTS

Next, we apply all the remedies on the Hilbert R-tree, the R*-tree and the RR*-tree. As the results are consistent in all three R-tree types, we only show the results of the Hilbert R-tree. Two data sets are used: uniform data and OpenStreetMap.

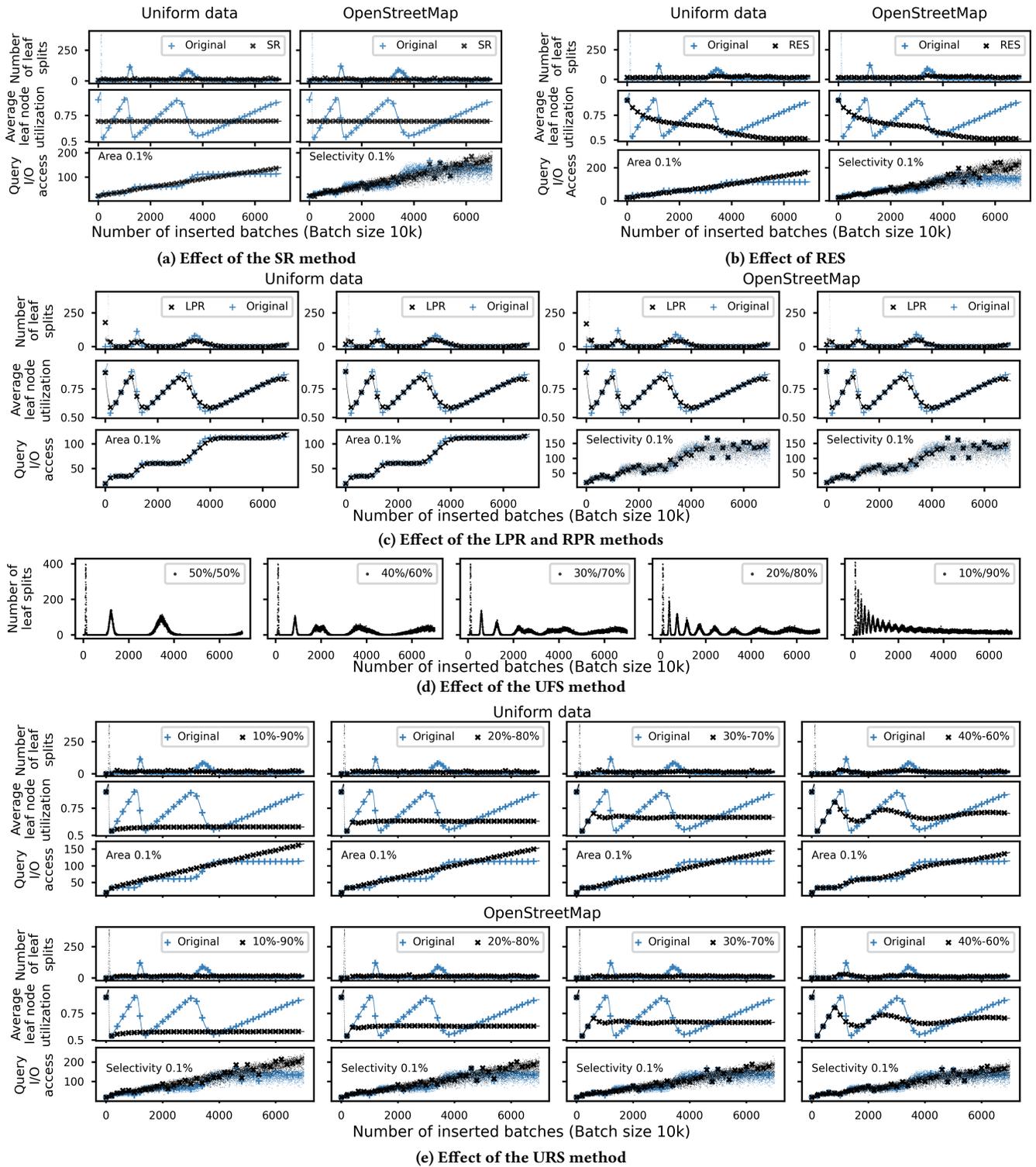

Figure 6: Waves of misery and query performance of the Hilbert R-tree with remedies. Both uniform data and `OpenStreetMap` data are used.

## 8.1 Effect of SR

We bulk-load the Hilbert R-tree using the new algorithm introduced in Section 7.1.1. The waves are completely eliminated with this new bulk-loaded R-tree (Figure 6a), proving the feasibility of SR. We observe the average leaf node utilization that remains stable at around 69%. This is consistent with the theoretical analysis. Without the waves of misery, query performance becomes stable and deterministic. We show one of the range sizes in Figure 6a and it does not show a staircase pattern as in the original Hilbert R-tree. With OpenStreetMap data, both waves of misery and average leaf node utilization are identical to the results from the case of the uniform data. The query result has a higher variation because data is skewed and the selectivity is an approximate number.

## 8.2 Effect of LPR and RPR

We evaluate the effects of LPR and RPR on Hilbert R-tree. From Figure 6c, observe that the peaks of waves of misery treated with remedies are decreased. The decrease is most effective for the first wave. Both LPR and RPR have their first wave immediately after batch insertion because portions of the nodes are nearly fully utilized. All of the three waves shown in one plot have almost equal height, and the batch number of each wave is almost identical to that of the original Hilbert R-tree. The average leaf node utilization is consistent with the waves of misery, where the oscillation range shrinks as the amplitude of the wave decreases. Also, the query performance is impacted. The increase between each plateau becomes slightly more flat with remedies. Both LPR and RPR show a similar effect in mitigating waves of misery, and the results are consistent for the two data sets.

## 8.3 Effect of URS

First, we investigate UFS and plot the waves of misery in Figure 6d. The waves still exist. If we compare the most skewed distribution (i.e., the 10%/90%) with the least one (i.e., the 50%/50%), we find that the waves are much denser in the skewed splits and converge faster. Also, we observe waves with two twin peaks in the 40%/60% and 30%/70% distributions. The reason is that after the split, each child with its fixed utilization forms its own wave, e.g., the wave from the 40% nodes and the wave from the 60% nodes. When the two rounds of waves collide, a wave with twin peak appears.

Next, we investigate the effect of URS in Figure 6e. The first wave still exists in all the tested ranges. However, the splits become stable after that. For the ranges 10%-90% and 20%-80%, the second and following waves are negligible. With a more narrowed range (30%-70% and 40%-60%), the second waves become clearer, but the heights of the waves are much lower compared to the original waves in all tested ranges. Average leaf node utilization correlates with the waves of misery. For range 10%-90%, the average leaf node utilization is around 57% after Batch 1000. This number is higher for range 20%-80% at around 63%. The 30%-70% range oscillates more but its stable average leaf node utilization is around 66%. The 40%-60% range oscillates the most but has the largest average leaf node utilization. Thus, there is a trade-off between waves of misery and leaf node utilization. For query performance, the R-trees with 10%-90% and 20%-80% ranges are worse in number of I/O accesses than the other two. Considering all the aspects above, we use the 30%-70% range for remedy comparison.

## 8.4 Effect of RES

With an insertion batch size of 10k, for RES, we test several split frequencies: one split every 400-900 insertions. The higher split frequency (400) means more splits are forced. Thus, more leaf nodes are created, and leaf nodes are less utilized. There is a clear trade-off between waves of misery and utilization. We choose 600 and 900 to be the split frequencies for the Hilbert R-tree and the R*-tree, respectively. We allow one overflow page per leaf node. As in Figure 6b, the waves of misery are almost eliminated using RES with few oscillations around Batch 4000. To compute the average leaf node utilization, we do not consider overflow pages because they are for transient storage only and will be inserted into the leaf nodes eventually. The query performance shows a smooth growth in number of I/O accesses, which is more deterministic than in the original Hilbert R-tree. The results are consistent for the two data sets.

## 8.5 Comparison of Methods

We compare the query performance of various remedies. In addition, we include the Hilbert R-tree with deferred split and the R*-tree with reinsertion in the comparison (Figure 7a and Figure 7b). DE refers to the Hilbert R-tree with deferred split and the R*-tree with reinsertion. Since LPR and RPR have similar performance in terms of waves of misery, we only include LPR in the study. We show the results of four range queries over uniform data.

In Figure 7a, of three larger range queries after Batch 6000, RES has the largest number of I/O accesses, with SR and URS next, and DE and LPR have the lowest I/O accesses. In Batches 4000-6000, LPR displays a plateau, with RES still having the largest number of I/O accesses, and DE the lowest. Before Batch 4000, the difference among RES, SR and URS becomes negligible with LPR showing sharp steps and taking as few I/O accesses as DE around Batches 1500 and 3000, and most I/O accesses around Batches 1700 and 4000. With the smallest range query (0.01%), RES takes fewer I/O accesses than SR and URS in Batches 2200-5300. After Batch 5300, RES is slightly worse than SR and DE uses the lowest number of I/O accesses among all. For the R*-tree in Figure 7b, the differences among remedies are smaller. URS has a slightly higher number of I/O accesses among all remedies in all query sizes, followed by RES and SR. Before Batch 1000, RES uses the smallest number of I/O accesses among all. With smallest query size (0.01%), LPR uses the smallest number of I/O accesses after around Batch 5800. In the remaining batches and query sizes, DE uses the lowest I/O counts.

Next, we compare the query performance of all the remedies with that for the original trees without remedies. We count the number of batches that the remedy outperforms the Hilbert R-tree out of 7000 batches in Table 1. Under all range sizes, DE has the largest number of batches that outperforms the Hilbert R-tree. With the smallest range size, DE, SR, RES and URS have better performance in more than half of the batches. For the R*-tree, we compute the equivalent numbers in Table 2. With very small range sizes (0.0001% and 0.001%), none of the remedies outperform the original R*-tree.

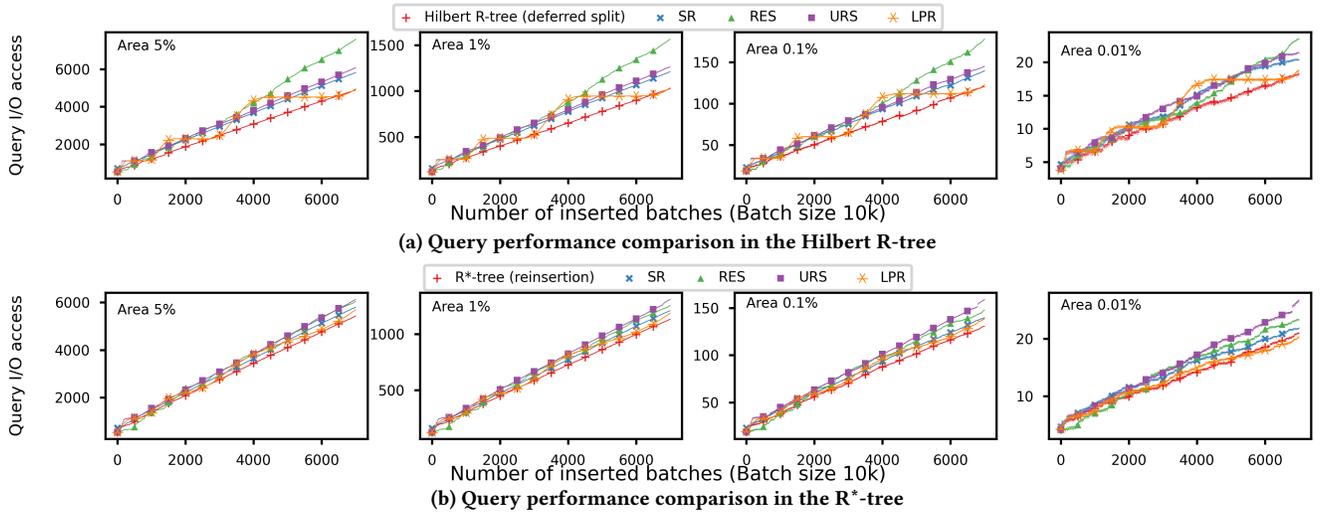

Figure 7: Query performance comparison in the Hilbert R-tree and the R*-tree

Table 1: Percentage that the remedy outperforms the Hilbert R-tree

| Number of batches | DE | SR | RES | URS | LPR |
|---|---|---|---|---|---|
| Query Range 5% | 75% | 41% | 28% | 31% | 20% |
| Query Range 1% | 77% | 43% | 31% | 39% | 59% |
| Query Range 0.1% | 79% | 46% | 35% | 39% | 73% |
| Query Range 0.01% | 82% | 39% | 47% | 35% | 67% |
| Query Range 0.001% | 82% | 57% | 54% | 32% | 50% |
| Query Range 0.0001% | 83% | 76% | 82% | 83% | 25% |

Table 2: Percentage that the remedy outperforms the R*-tree

| Number of batches | DE | SR | RES | URS | LPR |
|---|---|---|---|---|---|
| Query Range 5% | 86% | 41% | 33% | 18% | 51% |
| Query Range 1% | 90% | 45% | 40% | 18% | 58% |
| Query Range 0.1% | 84% | 45% | 38% | 16% | 64% |
| Query Range 0.01% | 34% | 0 | 16% | 0 | 29% |
| Query Range 0.001% | 3% | 0 | 6% | 0 | 0 |
| Query Range 0.0001% | 1% | 0 | 3% | 12% | 7% |

With larger range sizes, DE has a better performance over the original tree, while SR and RES are slightly worse, LPR has comparable performance, and URS is the lowest among all on average.

## 9 CONCLUDING REMARKS

This paper offers a new perspective for comparing R-trees using waves of misery. Waves of misery result in non-deterministic performance as clustered splits put extra stress on buffer management. Besides query performance, waves of misery also affect node utilization. Although the Linear and Quadratic R-trees are outperformed by the other R-tree variants in previous studies, the waves of misery of the Linear and Quadratic R-trees are negligible. Experiments show the uniqueness in their splitting algorithms that makes them not suffer from waves of misery. Several methods are proposed to alleviate waves of misery for the other R-tree types, and all of them have trade-offs in waves of misery and query performance.

- If insertion time is a tolerable issue, then deferred split is a good choice. The Hilbert R-tree with deferred split and the R*-tree with reinsertion have reasonable query performances, and only one or two waves of misery. If faster insertion time is needed, then SR, RES, URS, and PR can be good candidates.
- SR is most effective in eliminating waves of misery, and has comparable query performance. If tree rebuild is an option, SR and PR are good candidates. The advantages of PR are twofold: easy implementation and higher node utilization. However, PR is not as effective as SR in eliminating waves of misery.
- If tree rebuild is not an option, then we can use either RES or URS, where both have good query performance for queries of smallest size for the Hilbert R-tree. If we can tolerate some extra space for the index, then RES can be a good candidate because its regular split results in lower average leaf utilization. Otherwise, URS is a good choice.

## ACKNOWLEDGMENTS

Lu Xing thanks Dr. Kai Tang for the in-depth discussion on the splitting algorithms of the Linear and Quadratic R-trees. Walid G. Aref acknowledges the support of the National Science Foundation under Grant Numbers III-1815796 and IIS-1910216.